\documentclass[graybox]{svmult}

\usepackage{type1cm}        
\usepackage{makeidx}         
\usepackage{graphicx}        
\usepackage{multicol}        
\usepackage[bottom]{footmisc}
\usepackage{siunitx}
\usepackage[square, numbers, sort&compress]{natbib}

\usepackage{newtxtext}       %
\usepackage[varvw]{newtxmath}       
\newcommand{\dax}{\ensuremath{\mathrm{d}}}
\newcommand{\qax}{\ensuremath{\mathrm{q}}}
\newcommand{\str}{\ensuremath{\mathrm{s}}}

\usepackage{tikz}
\usepackage{pgfplots}
\pgfplotsset{compat=newest}
\usepgfplotslibrary{fillbetween}

\newcommand{\circlednumber}[1]{%
  \tikz[baseline=(char.base)]{\node[draw, circle, inner sep=1pt](char) {#1};}%
}

\makeindex             

\begin{document}

\title*{Characterization of permanent magnet synchronous machines based on semi-analytic model reduction for drive cycle analysis
}

\titlerunning{Semi-analytic characterization of PMSM}
\author{Leon Blumrich\orcidID{0009-0009-0648-0264}, Yves Burkhardt\orcidID{0009-0003-6913-9887} and Sebastian Sch\"ops\orcidID{0000-0001-9150-0219} }

\authorrunning{L. Blumrich, Y. Burkhardt and S. Sch\"ops}
\institute{Leon Blumrich \at Fachgebiet Elektrische Antriebssysteme, TU Darmstadt, Landgraf-Georg-Str. 4, Darmstadt\\
\email{leon.blumrich@eas.tu-darmstadt.de}
\and Yves Burkhardt \at Fachgebiet Elektrische Antriebssysteme, TU Darmstadt, Landgraf-Georg-Str. 4, Darmstadt\\
\email{yves.burkhardt@eas.tu-darmstadt.de}
\and Sebastian Sch\"ops \at Fachgebiet Computational Electromagnetics, TU Darmstadt, Schloßgartenstr. 8, Darmstadt\\
\email{sebastian.schoeps@tu-darmstadt.de}}

\maketitle

\abstract*{The characterization of an interior permanent magnet synchronous machine (IPMSM) requires numerical analysis of the nonlinear magnetic motor model in different load conditions. To obtain the case-specific best machine behavior, a strategy for the determination of stator input current amplitude and angle is employed for all possible load torques given a limited terminal current amplitude and DC bus voltage. Various losses are calculated using state of the art loss models. The electromagnetic performance of the electric machine is stored in lookup tables. These can then be used for the drive cycle analysis of the electric drive train in the design and optimization stages.\\
To avoid the use of a dedicated mesh generator in the numerical analysis, volumetric spline-based models are suggested. With this approach, the mesh can be generated directly from the Computer Aided Design geometry. This enables an automatic adaption of the grid following a geometry perturbation. With this the approximated solution is kept consistent over the different iterations of an overlying optimization, improving its convergence behavior.}

\abstract{The characterization of an interior permanent magnet synchronous machine (IPMSM) requires numerical analysis of the nonlinear magnetic motor model in different load conditions. To obtain the case-specific best machine behavior, a strategy for the determination of stator input current amplitude and angle is employed for all possible load torques given a limited terminal current amplitude and DC bus voltage. Various losses are calculated using state of the art loss models. The electromagnetic performance of the electric machine is stored in lookup tables. These can then be used for the drive cycle analysis of the electric drive train in the design and optimization stages.\\
To avoid the use of a dedicated mesh generator in the numerical analysis, volumetric spline-based models are suggested. With this approach, the mesh can be generated directly from the Computer Aided Design (CAD) geometry. This enables an automatic adaption of the grid following a geometry perturbation. With this the approximated solution is kept consistent over the different iterations of an overlying optimization, improving its convergence behavior.}

\section{Introduction}
\label{sec:1}

Permanent magnet synchronous machines (PMSM) have become the standard motor solution for high torque density applications within the last decades. Especially the electrification of passenger cars brought a new dynamic into the PMSM development. As this motor was and still is used in most electric vehicle (EV) drivetrain solutions, there is an increasing interest in optimization and detailed design \cite{fahimi_automotive_2024}. Typical optimization targets reach from performance characteristics like torque and power via drive cycle efficiency to cost driven Key Performance Indicators (KPIs) like overall material usage or magnet mass. Especially within the permanent magnets (PM), the content of heavy rare earth (HRE) material like Terbium or Dysprosium is a cost driver \cite{al-qarni_eliminating_2021} and therefore subject to minimization.\\

There are many more KPIs and design goals, e.g., there is a trend to increase motor rotational speeds to reduce the needed motor torque, which allows a reduction of the motor diameter and the overall volume resp. the material usage. In this context, today's EV motors in PMSM technology typically operate with a maximum rotational speed between \SI{15000}{\per\minute} and \SI{20000}{\per\minute}. Considering the rotor structural mechanics, the high rotational speed and the subsequently high circumferential speed ($>$ \SI[per-mode=symbol]{100}{\meter\per\second}) result in high mechanical stresses in the rotor lamination. The maximum admissible stress in the rotor lamination is a relevant limitation to the electromagnetic rotor design optimization. Therefore an efficient combination of electromagnetic and structural mechanical simulation is needed for a time efficient overall optimization. Today's commercial simulation tools offer coupled electromagnetic and structural mechanical simulations, typically creating different finite element (FE) models with a different mesh which is time consuming and needs to be synchronized at any design change in one of the domains. Geometrical details like radii in the edges of the rotor structure are highly relevant in terms of mechanical stress while there is only a minor influence on the electromagnetic behavior. Thus the level of geometric details especially in a parametric model deviates between electromagnetics and structural mechanics. A common geometry representation for both physical domains would be beneficial, allowing design variations without loosing the model validity in one of the domains.\\
\\
During the development of an automotive traction motor or entire drive train, the requirements and boundary conditions are typically subject to several changes like a change of direct voltage (DC voltage), exact torque-speed curve, installation space etc. For this reason, common simulation approaches use large scale design space exploration \cite{clauer_investigation_2023} (in some cases even full factorial) to provide sufficient data for a potential later change of the weighting factors in a target function avoiding further FE simulations. This is the motivation to develop a computational efficient electromagnetic and rotor structural mechanical simulation workflow, handling a large scale of independent 2D FE simulations. Furthermore, to foster fast and consistent design variations, a new spline-based model paradigm is advocated, e.g. \cite{Merkel_2021ab,Schops.2024}. \\
\\
In Section~\ref{sec:2}, the considered machine type and basic topology will be described in detail. Section~\ref{sec:3} presents our proposed semi-analytical simulation workflow, based on a minimum number of needed 2D FE simulations. We show approaches how to minimize the FE model dimensions and the number of needed simulation steps. In Section~\ref{sec:4}, the numerical definition and solution steps are explained. The spline-based model paradigm will be introduced which allows freeform simple geometry variations, even without additional meshing effort. The conclusion of the present paper is reached in Section~\ref{sec:5}.

\section{Electric Machine}
\label{sec:2}
\begin{figure}[b]
\sidecaption
\input{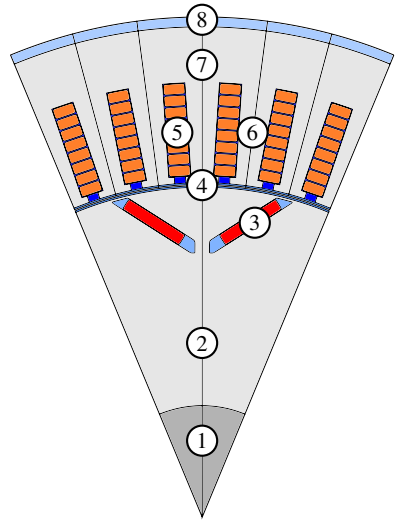}
\caption{Single pole model of the presented electric machine. The rotating part comprise the shaft \protect\circlednumber{1}, laminated rotor yoke \protect\circlednumber{2} and the magnets \protect\circlednumber{3} in their air pockets. Between rotor and stator is the air gap \protect\circlednumber{4}. The stator is comprised of the windings \protect\circlednumber{5} in the slots, seperated by the teeth \protect\circlednumber{6}, which are part of the laminated stator yoke \protect\circlednumber{7}. The air surrounding the stator back side is \protect\circlednumber{8}.}
\label{fig:machine} 
\end{figure}

This section is dedicated to the presentation of the electric machine (EM), accompanied by a comprehensive exposition of the conventional topological selections pertinent to automotive motors. Subsequent to this, a reference model is introduced.\\
The EM consists of rotating parts (rotor), comprising the shaft, rotor yoke, magnets, and air pockets. These are separated from the stationary part (stator) by the air gap. The stator consists of the housing, yoke, teeth, and slots with the winding system.\\
Today’s automotive EMs typically use topologies with a distributed stator winding system in hairpin technology and buried magnets, which are explained further in the following sections. The size and position of the components are commonly defined by parameters, and the general electromagnetic behavior can be analyzed in a cross-section perpendicular to the rotor’s rotational axis.\\
The example in this article features a typical automotive traction PMSM with V-shaped buried magnets and an eight-layered hairpin winding, as shown in Fig. \ref{fig:machine}.

\subsection{General Design}
General design parameters of the EM are the number of phases, the number of poles, skewing of the machine, and the materials and geometrical dimensions.\\
The majority of traction machines are three-phase machines. Higher phase numbers can be employed to increase the degrees of freedom in the machine design \cite{Levi.2008}. This can be used, for example, to improve the machines fault-tolerant operation or to increase the torque density (e.g. by injection of a third harmonic current \cite{Moeller.2022}). The multi-phase machine designs are seldom used in EV applications because of the additional complexity and cost regarding the power electronic inverter \cite{Salem.2019}.\\
The periodicity of the machine geometry is defined by its number of poles. This number is chosen as a compromise between a low magnetic flux per pole and low electrical frequency for a given rotational speed. A high number of poles reduces the magnetic flux per pole and thus saturation in the iron parts for a given yoke height. For a fixed rotational speed, the frequency of the winding currents and induced voltage increases with more poles. Higher frequencies lead to higher alternating current (AC) winding and iron losses, which need to be mitigated.\\
Skewing of the stator or rotor can be employed to reduce adverse effects, such as noise and vibrations, caused by the slotting of the stator \cite[Chapter 3]{Binder.2012}. Skewing requires an additional consideration in the 2D machine analysis due to its axial influence on the machine performance.\\
The design of the electric machine involves key choices related to material selection and geometric dimensions, which influence performance, efficiency, and manufacturability. These considerations apply to both the stator and rotor, each of which has specific requirements dictated by its function. Typical requirements regarding the volumetric dimensions of the EM are given by its application. In this case, they are specified by the automotive car makers (OEMs).

\subsection{Stator Design}
\label{sec:2_stator}
Design choices in the stator encompass the selection of materials, the properties of the winding system, and the geometrical characteristics of the components. \\
As mentioned in the previous paragraph, material selection and geometric dimensions play a crucial role in motor performance. The stator’s material choice and geometry affect electromagnetic properties, thermal management as well as the propagation of noise and vibrations.\\
The stator is slotted to accommodate the winding system, where typically the slots are semi-opened as a compromise between reducing stray paths for magnetic flux, minimizing harmonics in the magnetic field, and enabling manufacturability. The number of stator slots results from the desired number of poles and the selected number of slots per pole and phase $q$. Values of $q = 2$ or $q=3$ are often chosen to reduce the harmonic content in the airgap magnetic field.\\
The considered machine features profiled hairpin windings, which enable higher slot fill factors, improving torque and power densities compared to stranded round wire windings. The shorter end winding length mitigates thermal hotspots, caused by the poor thermal dissipation characteristics of the surrounding air and reduces losses due to magnetic stray flux \cite{Tola.2022}. Additionally, improved heat dissipation allows for higher current densities with the same cooling setup compared to round wire windings. \\
However, hairpin windings have notable drawbacks, including significant high-frequency conduction losses due to the skin and proximity effects, as well as reduced design flexibility due to limitations in the number of turns per coil and manufacturing constraints.
These challenges can be addressed by carefully selecting conductor layers and parallel current branches \cite{Venturini.2022}. In automotive traction applications, the number of hairpin layers tends to increase to counteract current displacement at higher speeds, with a typical number of eight layers \cite{Zou.2022} as adopted in this model (Fig. \ref{fig:machine}).
Despite these limitations, hairpin windings have become the standard in automotive applications due to their suitability for automated mass production.

\subsection{Rotor Design}
\label{sec:2_rotor}
The main design choices of the rotor include the number and arrangement of the PMs, the contour of its surface, and the material and geometry considerations introduced earlier. In particular, the selection of materials and the geometric arrangement affect magnetic and conductive losses, mechanical stability and torque production.\\
The rotor contains the PMs that excite the rotor magnetic field. The magnets are embedded in the rotor iron and angled relative to the circumferential coordinate to utilize flux concentration, enabling higher air gap flux densities compared to surface-mounted magnets. Additionally, the absence of magnets in the air gap -- where their low magnetic permeability ($\mu_{r, \text{pm}} \approx \mu_0$) would otherwise increase the effective air gap width -- enhances stator-rotor coupling and improves torque density. The integration in the rotor creates a difference in the reluctance in the d- and q-axis, which is called saliency, generating an additional reluctance torque.
Although electromagnetically unfavorable, a minimum bridge thickness between the magnets and the airgap is required to ensure structural integrity, especially at high rotational speeds.\\
Various magnet arrangements have been explored in literature \cite{Krings.2020}, though practical implementations are often constrained by the need to minimize manufacturing complexity. Common configurations include the plain single-V (as shown in Fig. \ref{fig:machine}), a $\nabla$ or $\Delta$ shape incorporating an additional bar magnet,  and the double-V design with two magnet layers. At later design stages, additional holes can be introduced into the rotor iron core to reduce the moment of inertia and improve the dynamic response or for axial rotor cooling purposes. These holes should not interfere with significant magnetic flux paths and are therefore typically positioned near the shaft.

\section{Modelling}
\label{sec:3}
This section introduces and motivates the machine characterization methodology. Unlike applications operating at a fixed supply voltage and frequency, optimizing the machine for a single operating point (OP) is not feasible for traction drives. In practice, a wide range of operating conditions must be considered. Although specific points in the torque-speed-profile  may be of special interest, the machine design must be based on the entire operating range.\\
Typically, test procedures based on drive cycles are used for the homologation of electric vehicles, which motivates their integration into the machine design stage. From the operating profile, the energy efficiency of the electric drive train and the range of the vehicle can be derived for a given machine design.\\ 
Drive cycle analysis of an electric machine requires models or assumptions for the longitudinal vehicle model, road and weather conditions, traction battery, inverter and gearbox. Such an operating profile consists of numerous OPs, typically sampled every second, and extends from 20–30 minutes and more, depending on the specific drive cycle. The most commonly used drive cycle is the Worldwide Harmonized Light-Duty Vehicle Test Cycle (WLTC).\\
Performing a full nonlinear time-stepping analysis over an entire drive cycle is computationally prohibitive. Alternative approaches include clustering recurring OPs or using a reduced order model of the machine  \cite{Carraro.2016}.
In this case, the nonlinear magnetic circuit of the machine is characterized a priori, and  the operational performance is analytically post-processed using the resulting lookup-table (LUT)-based model. This method assumes the piece-wise independence of all OPs, neglecting transient effects and thermal influences on the nonlinear magnetic circuit. The combination of preceding numerical evaluations and further analytical calculations defines the semi-analytical nature of the machine characterization process.

\subsection{Machine Characterization Process}
\label{sec:3_1}
\begin{figure}[t]
\sidecaption[t]
\input{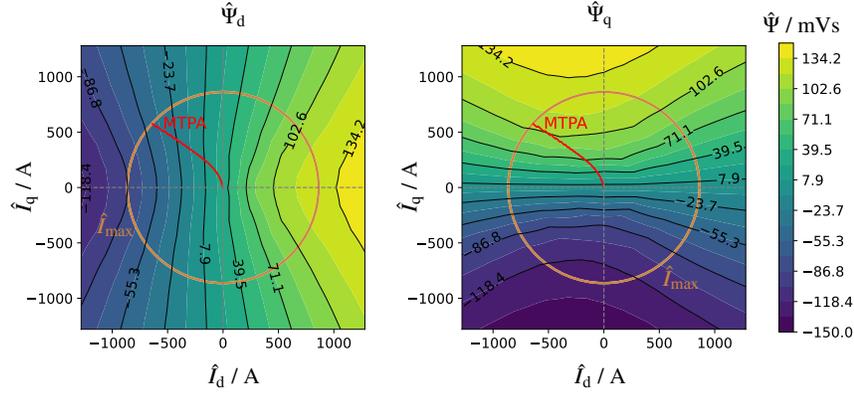}
\caption{$\hat{\Psi}_\dax$ and $\hat{\Psi}_\qax$ depending on the stator currents. The MTPA trajectory as explained in the following paragraph is shown in red. The current limit is depicted as a circle.}
\label{fig:flux_map} 
\end{figure}

The advantage of an a priori machine calculation is the reduced computational load for the drive-cycle evaluation and improved scalability to more complex testing procedures. The analytical post-processing of the LUT can be performed during the drive-cycle evaluation to account for updates in the DC-bus voltage or winding temperatures -- i.e., parameters that do not significantly affect the nonlinear magnetic circuit of the machine. Consequently, further FE analysis (FEA) is not required, keeping computation times low. Losses are calculated from the LUTs based on analytical models, which are further elaborated in this section. Since the exact value of power loss is significantly influenced by manufacturing effects and tolerances, a more accurate loss calculation using intricate material models may not be necessary \cite{Leuning.2021}.\\
Another advantage is that FEA needs to be conducted at only one speed, or at a few discrete speeds depending on the models used to account for speed-dependent losses. Torque production depends only on current, whereas the rotational speed determines the admissible stator current that satisfies the voltage limit. This calculation can be performed analytically, and losses can be scaled for different rotational speeds using their analytical models.\\
\\
The machine input is a sinusoidal stator current per phase, expressed as a complex phasor 
\begin{equation}
    \underline{I}_\str = \hat{I}_se^{-j\beta} 
\end{equation}
where $\hat{I}_s$ is the amplitude and $\beta$ the phase angle relative to the rotor position. The assumption of a purely sinusoidal excitation is known as fundamental analysis, where harmonic effects caused by the inverter operation are neglected.\\
The stator current can be decomposed into rotor-fixed coordinate system components using Park's transformation \cite[Chapter 9]{Binder.2012}. The direct-axis (d) and quadrature-axis (q) current components are given by: 
\begin{equation}
    \begin{bmatrix} I_\dax \\ I_\qax \end{bmatrix} = 
    \begin{bmatrix} \mathrm{cos}\;\beta \\ \mathrm{sin}\;\beta \end{bmatrix} \cdot I_\str,
\end{equation}
where $\beta$ is the angle from the d-axis to the current.
The nonlinear magnetic behavior is represented by the magnetic flux linkages per phase $\hat{\Psi}_\str (I_\dax, I_\qax)$, which are obtained from nonlinear analysis (described in Section~\ref{sec:4}). A typical flux linkage map, decomposed into d- and q-axis components, is shown in Fig. \ref{fig:flux_map}. The quantity $\hat{\Psi}_\str$ links the magnetic flux -- excited by stator and rotor currents or PMs -- to the phase winding, depending on its coil area and number of winding turns.

\subsection{Analytical Model}
\label{sec:3_2}
Electromagnetic performance characteristics can be calculated based on the flux maps. The electromagnetic torque of a three-phase motor is given by: 
\begin{equation}
    M = \frac{3 p}{2} (\hat{\Psi}_\dax\hat{I}_\qax - \hat{\Psi}_\qax\hat{I}_\dax),
    \label{eq:torque}
\end{equation}
where $p$ is the number of pole pairs. The same value of torque can be produced by different combinations of d- and q-current components. The desired combination is determined according to the current control principle. Typical objectives include achieving the lowest current amplitude $\hat{I}_\str$ respectively the maximum torque per ampere (MTPA) or minimizing the total losses. The most commonly used strategy is MTPA, which aims to reduce conduction losses in motor and inverter.\\
The stator voltage equation per phase is given by
\begin{eqnarray}
    \hat{U}_\dax &=& R_\str\hat{I}_\dax - \omega_\mathrm{el} (\hat{\Psi}_\qax + L_{\sigma, \mathrm{ew}}\hat{I}_\qax) \label{eq:u_1}\\
    \hat{U}_\qax &=& R_\str\hat{I}_\qax + \omega_\mathrm{el} (\hat{\Psi}_\dax + L_{\sigma, \mathrm{ew}}\hat{I}_\dax) \label{eq:u_2}\\
    \hat{U}_\str &=& \sqrt{\hat{U}_\dax^2 + \hat{U}_\qax^2} \label{eq:u_3}.
\end{eqnarray}
Stray flux in the end windings can be taken into account with the empirical, measured or estimated stray inductance $L_{\sigma, \mathrm{ew}}$.\\
The electrical frequency depends on the desired rotor speed $n$, for synchronous machines given by
\begin{equation}
   \omega_\mathrm{el} = 2\pi n p. 
\end{equation}
The second terms in the stator voltage equations \eqref{eq:u_1} and \eqref{eq:u_2} account for voltage induction due to the machine rotation and increase with rotational speed. A given maximum DC-bus voltage must not be exceeded in the stator phases $\hat{U}_\str \leq \frac{2}{\pi}U_\mathrm{dc}$, which limits the admissible operating range. Different overmodulation techniques exist to further increase the fundamental phase voltage \cite{Mahlfeld.2016}. Similarly, the maximum permissible phase current $\hat{I}_{\str, \text{max}}$ is constrained by the inverter switches and thermal limits of the winding insulation.

\subsection{Losses}
\label{sec:3_3}
During operation, power is dissipated and converted into heat in various machine components. This power does not contribute to the electromechanical power conversion and cannot be utilized for traction or recuperation, thus it is classified as loss. Losses can be categorized based on whether they vary with rotational speed $n$, load (i.e. $I_\str$), component or cause. The most significat loss categories include:
\begin{itemize}
    \item Conduction losses in the stator windings
    \item Core losses in the laminated stator and rotor iron cores, as well as the PMs
    \item Friction losses, caused by mechanical friction in the bearing and windage on the rotor surface.
\end{itemize}
Core losses are divided into three primary categories. Hysteresis losses, caused by the change of magnetic polarity in soft magnetic materials. Eddy current losses, as conduction losses caused by induction in conductive materials by time-varying magnetic fields. And finally, excess losses which account for additional dynamic and harmonic effects.\\ 
Additional losses include stray load losses, occurring in the non-active magnetic and conductive parts and harmonic losses caused by inverter-fed operation. These losses are typically small, require elaborate models or measurement setups, and are therefore not the focus of this study.

\subsubsection{Conduction losses}
For a three-phase machine, conduction losses are given by
\begin{equation}
    P_\mathrm{cu} = P_\mathrm{cu,dc} + P_\mathrm{cu,ac} = 3\cdot [R_{\str, dc}(T_\mathrm{cu}) + R_{\str, ad}(T_\mathrm{cu}, f)]\cdot I_\str^2
    \label{eq:p_cu}
\end{equation}
assuming symmetrical impedance of the three phases. These losses are divided into a frequency independent (DC) component, expressed by the DC resistance 
\begin{equation}
     R_{\str, \text{dc}} = \frac{l_\mathrm{c}}{\sigma_0 \cdot A_\mathrm{c}} (1 + \alpha(T - T_0)) \label{eq:r_dc}
\end{equation}
\eqref{eq:r_dc}, and a frequency dependent (AC) component
\begin{equation}
    R_{\str, \text{ac}} = R_{\str, \text{dc}}\cdot \frac{l_\mathrm{eff}}{l_\mathrm{c}} \cdot a \cdot f^{b}, \label{eq:r_ac}
\end{equation}
for which the resistance \eqref{eq:r_ac} is adapted to account for current displacement and proximity effects.

In \eqref{eq:r_dc}, $\sigma_0$ is the material-specific electric conductivity at reference temperature $T_0$ in \si[per-mode=symbol]{\siemens\per\metre} and $\alpha$ is the temperature coefficient in \si{\per\kelvin}. Geometrical parameters are the total conductor length $l_\mathrm{c}$ per phase and the conductor cross-sectional area $A_\mathrm{c}$. \\
Frequency-dependent losses are considered only in the winding region in the stator slots, where the effective conductor length is $l_\mathrm{eff}$. The empirical coefficients $a$ and $b$ are obtained by fitting results from machine simulations at different speeds. Further details about AC losses are provided in \cite{Hajji.2024}.

\subsubsection{Core losses}
The magnetic core losses are computed using Bertotti's formula, adapted for arbitrary periodic magnetic flux density waveforms \cite{Lin.2004}, i.e.,
\begin{equation}
    P_\mathrm{fe} = k_h\cdot B_\mathrm{max}^2\cdot f + \frac{k_c}{2\pi^2}\cdot \mathcal{\dot{B}}^2\cdot (\frac{f}{f_0})^2 + \frac{k_e}{8.76}\cdot \mathcal{\dot{B}}^{1.5}\cdot (\frac{f}{f_0})^{1.5}.
    \label{eq:p_fe}
\end{equation}
The material coefficients for hysteresis $k_h$, eddy currents $k_c$ and excess losses $k_e$ are required, which are typically either provided by the material suppliers or measured by the machine manufacturers.\\
To consider non-sinusoidal flux density waveforms based on the simulation at reference speed, the following integral is evaluated
\begin{equation}
    \mathcal{\dot{B}}^\alpha = \frac{1}{\tau_0} \int_0^{\tau_0} (\frac{\mathrm{d}B}{\mathrm{d}t})^\alpha dt
\end{equation} over one period with the reference frequency $f_0 = 1 / \tau_0$.

\subsubsection{Friction losses}
Friction losses from bearing resistance and windage on the rotor surface are highly dependent on the exact surface geometry and roughness, manufacturing tolerances and surrounding airflow conditions \cite{KargBulnes.2022}. Losses can be approximated using a simplified cubic model:
\begin{equation}
    P_\mathrm{fr} = k_\mathrm{r1} f + k_\mathrm{r2} f^2 + k_\mathrm{r3} f^3,
\end{equation}
where the coefficients are extrapolated from measurement data or determined empirically.

\subsubsection{Efficiency}
\begin{figure}[t]
\sidecaption
\input{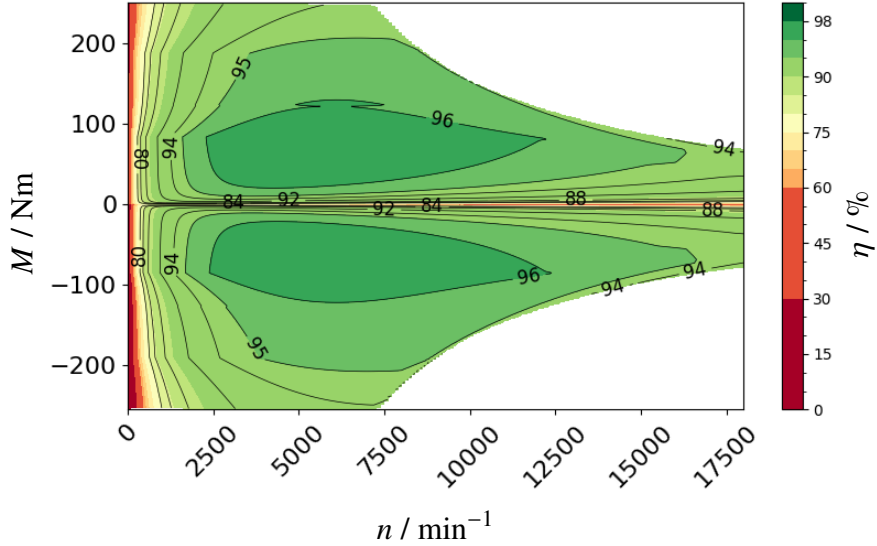}
\caption{Efficiency plot of an 8-pole IPMSM.}
\label{fig:eta} 
\end{figure}

Given the total losses in all OPs 
\begin{equation}
    P_\mathrm{loss} = P_\mathrm{cu} + P_\mathrm{fe} + P_\mathrm{fr},    
\end{equation}
the efficiency can be computed separately for motor and generator operation.
\begin{eqnarray}
    \eta_\mathrm{mot} &= \frac{P_\mathrm{m, out}}{P_\mathrm{el, in}} &= \frac{2\pi n M}{2\pi n M + P_\mathrm{loss}}\\
    \eta_\mathrm{gen} &= \frac{P_\mathrm{el, out}}{P_\mathrm{m, in}} &= \frac{2\pi n M + P_\mathrm{loss}}{2\pi n M},
\end{eqnarray}
where, $M, P_\mathrm{m} > 0$ for motor operation and $M, P_\mathrm{m} < 0$ for generator operation, using the passive sign convention. The efficiency map (Fig. \ref{fig:eta}) provides a comprehensive overview of the machine efficiency across its entire operating range. At low speeds, the DC conduction losses dominate. At high speed and high torque the AC conduction losses dominate with significant impact of the core losses at high speeds.

\subsection{Reduction of the workload}
\label{sec:3_4}
To minimize computational effort and reduce calculation time for a given design variant, the model complexity is reduced by exploiting symmetries in both time and space, thereby limiting the number of necessary calculations. Additionally, computational methods are carefully selected and adapted to suit the specific problem at hand. This section focuses on the first approach, detailing how computational workload is reduced through strategic simplifications.\\

As discussed in Section~\ref{sec:3_1}, numerical analysis is performed on the given design geometry for a discrete set of input currents $I_\dax, I_\qax$. From these results, the machine performance is evaluated analytically across the full range of torque and speed. Each numerical computation employs a quasi-stationary time-stepping analysis to obtain solutions for different rotor positions.

\subsubsection{Reduction of the geometrical model}
\begin{figure}[t]
\sidecaption[t]
\input{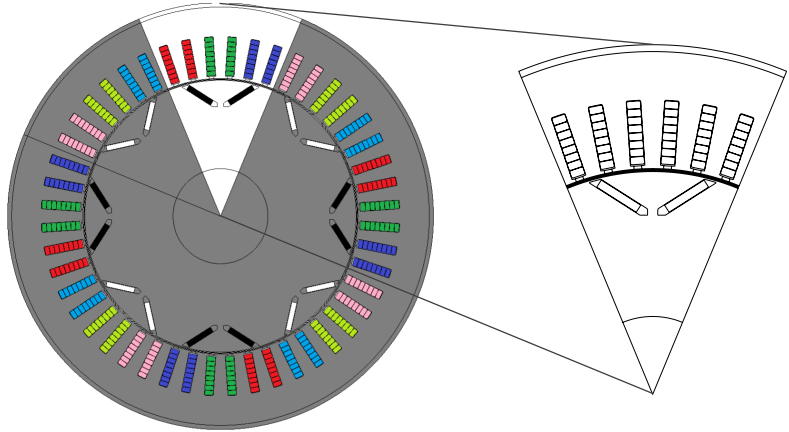}
\caption{Symmetries in the full machine model (left) allow the reduction of the calculation domain to a single pole. For integer-slot windings, shown here with two slots per pole and phase ($q = 2$), the magnetic field repeats anti-periodically for each subsequent pole.}
\label{fig:geo_reduct} 
\end{figure}

As introduced in Section~\ref{sec:2}, the machine model is reduced to 2D to significantly reduce the computational domain. This reduction is justified by the dominance of the radial and tangential components of the electromagnetic field, and reasonable accuracy can be maintained by adapting the model and material properties to account for the effects neglected due to the omission of the full three dimensional (3D) topology~\cite{Steentjes.2015}. These effects primarily include magnetic stray fields in the end regions and axial eddy current paths in conductive materials. To achieve maximum accuracy, 3D analysis is typically employed during the final design stages~\cite{Rosu.2017}.\\
In traction motors, the rotor is often skewed to reduce the effects of stator slotting on the torque ripple. These effects are incorporated into the 2D model by using a multi-slice method \cite{Williamson.1995}. Additionally, in multi-pole topologies, the geometry of the machine repeats periodically, for the considered types after every pole. The polarity of the winding currents and the PM magnetization alternates with each pole, enabling a reduction of the electromagnetic calculation domain to a single pole. This symmetry is illustrated in Fig. \ref{fig:geo_reduct}.\\
While this reduction is applicable to most machine topologies, some fractional-slot winding configurations require a model larger than a single pole due to a lack of winding periodicity. In such cases, the lowest-order field harmonic spans a wavelength exceeding one pole pair and defines the minimum symmetry unit size.\\
Furthermore, a full model of the machine is necessary to analyze the effects of rotor eccentricity, where the magneto-motive force (MMF) can no longer be assumed to be anti-periodic across poles.

\subsubsection{Reduction of the current sweep range}
Spatial symmetries can be used to reduce the domain of the complex plane of input currents $\underline{I}_\str$ required for the machine analysis. The flux linkage is symmetric around the d-axis, as expressed in \begin{equation}
    \Psi (\pi + \beta) = \Psi (\pi - \beta),
    \label{eq:psi_sym1}
\end{equation} meaning that it is sufficient to simulate only cases with positive q-current, i.e., $\beta < \pi$. This symmetry condition is broken when rotor eccentricity or significant deviations due to manufacturing tolerances are considered~\cite{Jaeger.2018}.\\
\\
Furthermore, the torque expression \eqref{eq:torque} can be rewritten as
\begin{equation}
    M = \frac{3p}{2} (\Psi_\mathrm{PM}\hat{I}_\qax + (L_\dax - L_\qax)\hat{I}_\dax\hat{I}_\qax), \label{eq:torque_2}
\end{equation}
using the flux linkages as given by
\begin{eqnarray}
    \Psi_\dax &=& \Psi_\mathrm{PM} + L_\dax \hat{I}_\dax \label{eq:psi_d}\\
    \Psi_\qax &=& L_\qax \hat{I}_\qax. \label{eq:psi_q}
\end{eqnarray} 
Cross-inductance terms are omitted due to their relatively minor effect.\\
\\
For an IPMSM, the d-axis corresponds to the flux path across the magnets. Due to the magnetic properties of the PMs being similar to air (permeability $\mu_\mathrm{PM} \approx \mu_0$), the d-axis inductance is lower than the q-axis inductance, i.e., $L_\dax - L_\qax < 0$. From \eqref{eq:torque_2}, it is evident that a negative d-current is required to increase the torque by use of reluctance. Thus, in non-faulty operation, it is sufficient to analyze negative d-axis current components, restricting the current angle to
\begin{equation}
    \frac{\pi}{2} < \beta \leq \pi.
\end{equation}
When using a multi-slice model for an EM with a skewed rotor or stator, and additional increment of the current angle must be considered since the effective current angle varies across slices.\\
The step widths for input current amplitude and angle should be chosen based on the machine's saturation behavior, ensuring that interpolation errors remain below an acceptable threshold and avoiding aliasing artifacts that deteriorate the solution accuracy \cite{Jaeger.2018}.

\subsubsection{Range of rotational angles}
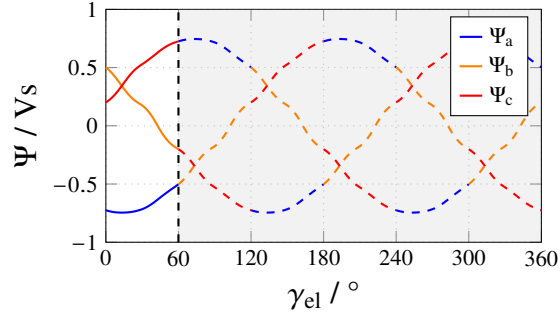
\begin{figure}[t]
\sidecaption[t]
\begin{tikzpicture}
    \begin{axis}[
        width=0.63\textwidth,
        height=0.4\textwidth,
        xlabel={\large $\gamma_\text{el}$ / °},
        ylabel={\large $\Psi$ / Vs},
        xmin=0, xmax=360,
        ymin=-1, ymax=1,
        xtick={0,60,120,180,240,300,360},
        ytick={-1,-0.5,0,0.5,1},
        legend pos=north east,
        legend style={font=\small},
        legend image post style={xscale=0.5},
        ticklabel style={font=\footnotesize},
        grid=both,
        grid style={dotted,gray!50}
    ]
        \addplot[blue, thick] table[x index=0, y index=1, restrict x to domain=0:60] {figures/flux_comp.txt};
        \addlegendentry{$\Psi_\text{a}$};
        \addplot[orange, thick] table[x index=0, y index=2, restrict x to domain=0:60] {figures/flux_comp.txt};
        \addlegendentry{$\Psi_\text{b}$};
        \addplot[red, thick] table[x index=0, y index=3, restrict x to domain=0:60] {figures/flux_comp.txt};
        \addlegendentry{$\Psi_\text{c}$};

        \addplot[orange, dashed, thick] table[x index=0, y index=1, restrict x to domain=60:120] {figures/flux_comp.txt};
        \addplot[red, dashed, thick] table[x index=0, y index=2, restrict x to domain=60:120] {figures/flux_comp.txt};
        \addplot[blue, dashed, thick] table[x index=0, y index=3, restrict x to domain=60:120] {figures/flux_comp.txt};

         \addplot[red, dashed, thick] table[x index=0, y index=1, restrict x to domain=120:180] {figures/flux_comp.txt};
        \addplot[blue, dashed, thick] table[x index=0, y index=2, restrict x to domain=120:180] {figures/flux_comp.txt};
        \addplot[orange, dashed, thick] table[x index=0, y index=3, restrict x to domain=120:180] {figures/flux_comp.txt};

         \addplot[blue, dashed, thick] table[x index=0, y index=1, restrict x to domain=180:240] {figures/flux_comp.txt};
        \addplot[orange, dashed, thick] table[x index=0, y index=2, restrict x to domain=180:240] {figures/flux_comp.txt};
        \addplot[red, dashed, thick] table[x index=0, y index=3, restrict x to domain=180:240] {figures/flux_comp.txt};

         \addplot[orange, dashed, thick] table[x index=0, y index=1, restrict x to domain=240:300] {figures/flux_comp.txt};
        \addplot[red, dashed, thick] table[x index=0, y index=2, restrict x to domain=240:300] {figures/flux_comp.txt};
        \addplot[blue, dashed, thick] table[x index=0, y index=3, restrict x to domain=240:300] {figures/flux_comp.txt};

         \addplot[red, dashed, thick] table[x index=0, y index=1, restrict x to domain=300:360] {figures/flux_comp.txt};
        \addplot[blue, dashed, thick] table[x index=0, y index=2, restrict x to domain=300:360] {figures/flux_comp.txt};
        \addplot[orange, dashed, thick] table[x index=0, y index=3, restrict x to domain=300:360] {figures/flux_comp.txt};

        \draw[dashed, thick, black] (axis cs:60,-1) -- (axis cs:60,1);
   
        \path[name path=top] (axis cs:60,1) -- (axis cs:360,1);
        \path[name path=bottom] (axis cs:60,-1) -- (axis cs:360,-1);
        \addplot[gray!50, opacity=0.2] fill between[of=top and bottom];

    \end{axis}
\end{tikzpicture}
\caption{Reconstruction of $\Psi$ over one electrical period from the simulated first sixth of a period by inversion and phase shifting. The solid lines show the simulated phase values , while the dashed lines indicate their continuation over the full period.}
\label{fig:psi_decomp} 
\end{figure}

As the rotor position changes, the magnetic paths and resulting flux linkages between the stator winding system and the magnetic field also change. Ideally, the electromagnetic torque of the motor should remain constant during the operation. However, due to slotting effects and the nonlinear behavior of the core, the torque varies with rotor position.\\
In a symmetric winding system, the flux linkage is anti-periodic over each pole. Additionally, the flux linkage of each phase is shifted by \SI{120}{\degree} electrical. Consequently, for a full pole pair (i.e. $\gamma_\text{el} = $ \SI{360}{\degree}), the phase flux linkages $\Psi_\text{a}$, $\Psi_\text{b}$ and $\Psi_\text{c}$ can be reconstructed from the first $\gamma_\mathrm{el} = $ \SI{60}{\degree} by inversion and phase shifting. This process is illustrated in Fig. \ref{fig:psi_decomp}.
The d- and q-axis flux linkages, $\Psi_\dax$ and  $\Psi_\qax$, are then derived from the phase components using Park's transformation.\\
Since a simulation of the first $\gamma_\mathrm{el} = $ \SI{60}{\degree} is sufficient, the corresponding mechanical rotation angle for simulation is
\begin{equation} 
    \gamma_\mathrm{m} = \frac{\gamma_\mathrm{el}}{p}. 
\end{equation}
For typical automotive traction machines, where the number of pole pairs is $p = 3$ or $p = 4$, the necessary mechanical rotation is $\gamma_\mathrm{m} = $ \SI{20}{\degree} and \SI{15}{\degree}, respectively. The step width for the rotational angle should match the discretization of the mesh elements at the boundary between moving and stationary regions to prevent numerical oscillations during the evaluation \cite{Ferreira.2002}.

\section{Numerical Analysis}
\label{sec:4}
The nonlinear characteristics of the magnetic sheet material necessitate a numerical analysis of the magnetic field distribution within the machine. Section~\ref{sec:3_1} outlines how FEA is employed in the context of the semi-analytic machine characterization process.\\
In FEA, the computational domain $\Omega$ is discretized into a mesh consisting of a finite number of geometrically small elements, over which the governing equations are solved in a weak sense. The solution is sought as a linear combination of weighted basis functions and thus an approximation of the true solution.\\
This following section particularly addresses spline-based geometry descriptions since this approach can eliminate the need for conventional mesh generators, thereby reducing computational effort and mitigating mesh sensitivity issues that can arise during the optimization stage due to slight variations in the geometry.

\subsection{Geometry description}
Electric machine models are frequently evaluated multiple times in various applications, such as shape optimization or uncertainty quantification \cite{Offermann_2015aa,Kuci_2017aa,Galetzka_2019aa}. 
In such cases, the geometry often requires local modifications, which require remeshing or cumbersome mesh morphing algorithms in conventional FEA workflows \cite{Dokken.2018}.
To address these challenges, we advocate a volumetric spline-based geometry description, which is still able to represent conic sections exactly while providing more design freedom at the same time \cite{Cohen_2001aa}. An advantage of this description is the ability to easily modify the geometry by adjusting the control points of the splines. When local modifications of the geometry are required, such a geometry description allows for automated mesh deformation. This can be directly incorporated into tools implementing the spline-based isogeometric version of the finite element method \cite{Merkel_2021ab,Komann_2024aa} or in conventional FEA tools if a mesh is extracted \cite{Schops.2024}. In the latter case, existing workflows can still be used with a spline based geometry description, e.g., the open-source tools Gmsh/GetDP can import such geometries \cite{Geuzaine_2009aa,Geuzaine_2009ab, Geuzaine.2007}. This will be pursued in the following.

\subsection{High fidelity machine model}
In the following, we assume that the relevant parts of the computational domain $\Omega$ and its subdomains, e.g. corresponding to coils, rotor, stator and PMs are defined using a spline-based geometry representation.\\
For a given stator current $\underline{\hat{I}}_\str$, the magneto-quasistatic problem is formulated as:
\begin{align}
        \nabla \times (\nu\nabla\times\boldsymbol{A}) + \sigma\frac{\partial \boldsymbol{A}}{\partial t} + \sigma \nabla\phi &= \boldsymbol{J}_{\text{s}} + \nabla\times (\nu \boldsymbol{B}_\mathrm{r}) && \text{in  } \Omega \label{eq:pde}\\
        \boldsymbol{A} &= 0 && \text{on  } \Gamma_0 \\
        \boldsymbol{A} (r, \pi-\varphi_0) &= - \boldsymbol{A} (r, \pi + \varphi_0) && \text{on  } \Gamma_\mathrm{sym}
\end{align}
where $\nu = 1 / \mu$ is the nonlinear magnetic reluctivity, $\sigma$ is the electrical conductivity and $\boldsymbol{B}_\mathrm{r}$ is the remanent magnetic flux density of the PMs. The magnetic vector potential and electrical scalar potentials are introduced:
\begin{align}
    \boldsymbol{B} &= \nabla\times \boldsymbol{A} & 
    \boldsymbol{E} &=-\frac{\partial\boldsymbol{A}}{\partial t}-\nabla\phi.
    \label{eq:ansatz}
\end{align}
The sinusoidal source current density in the windings
\begin{equation}
    J_\str = \hat{I}_\str\cdot \frac{N}{a A_\mathrm{c}}\cdot \mathrm{sin} (2\pi f t + \varphi),
    \label{eq:src_density}
\end{equation}
which depends on phase and polarity, is derived from $\underline{\hat{I}}_\str$ using the total conductor cross-section per slot $A_\mathrm{c}$, the number of winding layers per slot $N$ and the number of parallel paths $a$. In \eqref{eq:src_density}, $f$ denotes the excitation frequency and $\varphi$ is the phase shift angle.\\
The computational domain is discretized, the problem \eqref{eq:pde} is transformed into its weak formulation and discretized in time and space. The solution is approximated using the Ritz-Galerkin method, and the resulting nonlinear system of equations is solved using GetDP.\\
\begin{figure}[t]
    \sidecaption[t]
    \input{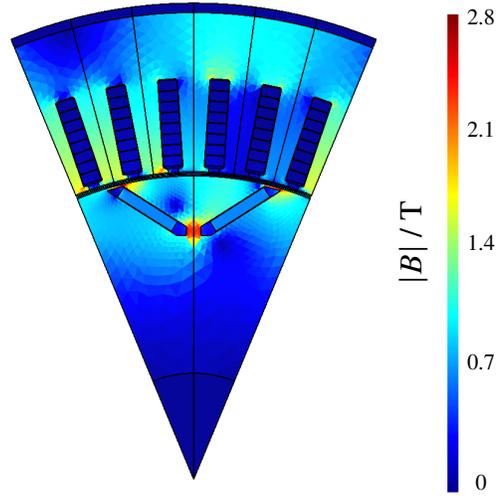}
    \caption{Magnetic flux density distribution calculated from the solution of \eqref{eq:pde} for an input current amplitude of \SI{300}{\ampere} and current angle of \SI{126}{\degree} at a rotational speed of \SI{6500}{\per\minute}. The low relative permeability of the windings compared to the surrounding iron yoke is evident. High flux densitiy saturates the rotor magnet bridges, increasing the reluctance of this flux path.}
    \label{fig:plot_b}
\end{figure}
The absolute magnetic flux density, derived from the field solution $\boldsymbol{A}$ in \eqref{eq:ansatz}, is shown for a fix time in Fig. \ref{fig:plot_b}. The calculation is performed for an input current of $\underline{I}_\str=\SI{300}{\ampere} \cdot e^{-j\cdot\SI{126}{\degree}}$, corresponding to a current density of approximately \SI[per-mode=symbol]{5}{\ampere\per\square\milli\metre} for the presented machine design, at a speed of \SI{6500}{\per\minute}.\\ 
The magnetic flux linkage per phase $\hat{\Psi}_\str$ is recovered from $\boldsymbol{A}$ according to \eqref{eq:fluxlinkage} by integrating the vector potential over the coil surface:
\begin{equation}
    \hat{\Psi}_\str = \int_{\boldsymbol{S}_\str} \frac{l_\mathrm{fe} N}{a A_\mathrm{c}} \boldsymbol{A}\cdot \mathrm{d}\boldsymbol{S}_\str
    \label{eq:fluxlinkage}
\end{equation}
The phase components are then transformed to the d-q-coordinate system by applying Park's transformation.

\subsection{Parallelization}
For the machine characterization, as introduced in Section~\ref{sec:3}, the stator current sweep is performed by independently solving \eqref{eq:pde} for each working point. This independence allows for parallel execution, maximizing computational efficiency by leveraging the available hardware resources. Parallelization is particularily beneficial for large-scale design studies, where multiple design variants are evaluated to meet different customer specification, as introduced in Section~\ref{sec:1}. The significance of parallelization is illustrated in the following simple example.

\begin{trailer}{Calculation example}
    Consider the evaluation of $n_\text{d} = 1000$ design variants, where each design requires $n_\text{i} = 20$ current amplitude steps and $n_\beta = 12$ current angle steps. Since the design is analyzed only once for $\hat{I}_\str = 0$, the total number of FE simulation jobs per design is
    \begin{equation*}
        n_\text{j} = (n_\text{i} - 1) \cdot n_\beta + 1 = 229.
    \end{equation*}
    Assuming that each simulation takes $t_\text{j} = $ \SI{35}{\second}, the total computation time for a serial execution would be:
    \begin{equation*}
        t_\text{serial} = n_\text{d}\cdot n_\text{j}\cdot t_\text{j} \; \hat{\approx}\; \SI{2226}{\hour} \;\SI{23}{\minute}.
    \end{equation*}
    However, by utilizing parallel computation on 64 cores, the total computation time is approximately reduced to \SI{34}{\hour} \SI{47}{\minute}, which is a reduction by 98~\%.\\
    The reduction of the simulation time from over 93 days down to under 2 days illustrates the significance of parallel processing in the efficient exploration of large-scale design spaces in EM design and optimization.
\end{trailer}

\section{Conclusion}
\label{sec:5}
This paper presented a semi-analytic workflow for the efficient characterization of a large number of electromagnetic designs for a PMSM to be used in a drive cycle analysis. By reducing the size of the simulation model, exploiting symmetry conditions, and minimizing the number of simulation time steps to 1/6 of an electric period, and parallelizing of the computation, computationally efficient 2D FE simulations could be achieved. All typical performance maps and KPIs were generated by an analytical workflow based on the basic FE motor characterization. To further accelerate the machine characterization, FE analysis using a spline-based motor geometry description was introduced. This avoids the classic FE meshing effort for every design variation while ensuring a consistent geometry representation for electromagnetic and structural mechanical evaluation. On this basis, time efficient large scale design space explorations can be performed as well as a detailed shape optimization for the final design.

\begin{acknowledgement}
This work is supported by the joint DFG/FWF Collaborative Research Centre CREATOR (DFG: Project-ID 492661287/TRR 361; FWF: 10.55776/F90    ) at TU Darmstadt, TU Graz and JKU Linz.
\end{acknowledgement}

\bibliographystyle{spmpsci}
\bibliography{references,bibtex}

\end{document}